\documentclass[sigconf]{acmart}

\AtBeginDocument{%
  }
\setcopyright{acmcopyright}
\copyrightyear{2026}
\acmYear{2026}
\acmDOI{XXXXXXX.XXXXXXX}

\acmConference[SIGIR VulGen '26]{International Workshop on Vulnerabilities in Generative Systems for Information Retrieval}{July 24, 2026}{Melbourne, Australia}

\setcopyright{acmcopyright} 

\acmISBN{978-1-4503-XXXX-X/2018/06}

\begin{document}

\title{Secure Coding Drift in LLM-Assisted Post-Quantum Cryptography Development: A Gamified Fix}


\author{R.D.N. Shakya}
\affiliation{%
  \institution{University of Moratuwa}
   \country{Sri-Lanka}
   \city{Moratuwa}}
\email{shakyardn.26@uom.lk}

\author{C.P. Wijesiriwardana}
\affiliation{%
  \institution{University of Moratuwa}
  \country{Sri-Lanka}
  \city{Moratuwa}}
\email{chaman@uom.lk}

\author{S.M.Vidanagamachchi}
\affiliation{%
  \institution{University of Ruhuna}
  \country{Sri-Lanka}
  \city{Matara}}
\email{smv@dcs.ruh.ac.lk}

\author{Nalin A.G. Arachchilage}
\affiliation{%
  \institution{RMIT University}
  \country{Australia}
  \city{Melbourne}}
\email{nalin.arachchilage@rmit.edu.au}

\renewcommand{\shortauthors}{R.D.N. Shakya et al.}

\begin{abstract}
The transition to Post-Quantum Cryptography (PQC) introduces considerable implementation complexity, requiring strict adherence to constant-time execution, side-channel resistance, and precise parametrisation. Simultaneously, large language models (LLMs) are heavily embedded in software development workflows, including cryptographic engineering. While LLMs improve productivity, evidence shows that they frequently generate insecure or suboptimal code, particularly in security-critical domains. This paper introduces Secure Coding Drift in PQC, a novel socio-technical vulnerability model capturing the gradual degradation of secure coding practices due to sustained reliance on LLM-generated code. Unlike prior work that focuses on static vulnerabilities, we conceptualise security risk as a longitudinal behavioural phenomenon rising from human–AI interaction. To mitigate this, we propose a gamified, LLM-augmented secure coding framework that embeds adversarial evaluation, behavioural feedback, and security scoring into development workflows. Our approach reframes LLMs from passive assistants into active security co-pilots, contributing toward safer PQC implementation in AI-mediated environments.
\end{abstract}



\received{20 February 2007}
\received[revised]{12 March 2009}
\received[accepted]{5 June 2009}

\maketitle

\section{Introduction and Background}

The rapid advancement of quantum computing threatens widely used public-key cryptographic systems such as RSA and elliptic curve cryptography, as quantum algorithms like Shor’s algorithm can solve their underlying mathematical problems in polynomial time \cite{bernstein_post-quantum_2017, aydeger_towards_2024}. In response, Post-Quantum Cryptography (PQC) has emerged as the primary defence against quantum attacks, with the National Institute of Standards and Technology (NIST) \cite{nist_pqc_project} leading the standardisation of quantum-resistant algorithms. However, migrating to PQC is not a simple replacement of algorithms. It requires integration into complex systems while maintaining security, performance, and interoperability \cite{aydeger_towards_2024}. PQC implementations also introduce engineering challenges such as large key sizes, performance overhead, and strict parameter control \cite{hekkala_implementing_2023}. More critically, they must satisfy strict security requirements, including constant-time execution and resistance to side-channel attacks. Even minor implementation errors can break theoretical security guarantees and lead to practical attacks \cite{hekkala_implementing_2023, toruan2026security}. Therefore, secure coding is not merely desirable in PQC software development; it is a fundamental requirement for preserving cryptographic security. 

At the same time, software engineering is shifting toward AI-native development methods, such as \textit{vibe coding} \cite{karpathy2025vibecoding}. In this approach, developers rely heavily on large language models (LLMs) for code generation and debugging \cite{HAQUE2025100204}. While this improves productivity, it introduces serious security risks. Studies show that LLM-generated code often violates secure coding practices and may contain vulnerabilities even when functionally correct \cite{asleep, Rami}. This is especially problematic in cryptographic contexts \cite{asleep}. For example, empirical studies on tools like GitHub Copilot show systematic failures to meet secure coding standards, especially in cryptographic contexts\cite{asleep}. These risks are further amplified by prompt injection, data poisoning, and other adversarial attacks on LLM pipelines \cite{BERINI2026104241}. In addition, developers often accept AI-generated code with limited verification, especially when the output appears syntactically correct or authoritative \cite{AI_worst, Insecure}. Over time, this over-reliance can reduce developers’ critical evaluation skills and internal security reasoning, creating \textit{cognitive debt}: a gradual erosion of security-related knowledge, behaviours and vigilance caused by habitual dependence on AI-generated solutions \cite{devDebt1, devDebt2, CogDebt1, cogDebt2}.

Existing research typically treats LLM-related security issues as static problems in generated code \cite{asleep, AI_worst, 11269659}. While this perspective is valuable, it overlooks the dynamic behavioural changes that emerge through prolonged interaction between developers and AI assistants \cite{devDebt1, devDebt2}. In practice, the security impact of LLM assistance is not only static but cumulative. Repeated exposure to auto-generated solutions can progressively reshape how developers think, reason, and validate code. This phenomenon motivates the concept of \textit{Secure Coding Drift} (SCD) \cite{apiiro_security_drift}, which describes the gradual degradation of secure coding behaviour over time due to increasing reliance on AI tools. Developers may perform less verification, weaker threat analysis, reduced awareness of edge cases, and accept insecure patterns more easily \cite{GLAS2026104842, apiiro_security_drift}. Rather than appearing as an immediate failure, drift manifests incrementally through behavioural adaptation, making it difficult to detect until insecure habits become normalised. Thus, the core risk is not only insecure generated code but also the long-term weakening of developers’ security mindset.
These issues are more critical in PQC development. Unlike traditional software development, PQC implementation requires deep understanding of subtle cryptographic properties that are often non-intuitive and highly sensitive to small coding decisions. A developer relying on vibe coding may receive code that appears mathematically valid and functional yet silently violates critical security assumptions such as constant-time execution or side-channel resistance. Since many PQC vulnerabilities are not visible through functional testing, insecure implementations may remain undetected until exploited \cite{hekkala_implementing_2023, bagirovs_applications_2024}. Furthermore, if developers increasingly trust LLM outputs without rigorous validation, secure coding drift can compound PQC implementation risks. Cognitive debt reduces the possibility that developers will question hidden cryptographic assumptions, while LLM vulnerabilities such as poisoned training data or adversarial prompting may further propagate insecure implementation patterns \cite{devDebt1, devDebt2}. Consequently, the intersection of vibe coding and PQC creates a high-risk environment in which both machine-generated vulnerabilities and human behavioural degradation jointly threaten software security.

Despite growing research on AI-assisted coding and PQC, limited work examines their long-term behavioural interaction. Most mitigation approaches focus on static solutions such as vulnerability detection tools rather than developer behaviour. To address this gap, we introduce \textit{Secure Coding Drift in PQC (SCD-PQC)}, a behavioural and longitudinal model that explains how secure coding practices degrade in LLM-assisted PQC development environments. Building on this model, we propose a mitigation framework grounded in gamification and continuous feedback to reinforce secure coding awareness, encourage active verification of AI-generated outputs, and sustain long-term secure development habits. By shifting the focus from one-time vulnerability detection to behavioural resilience, this work aims to support safer AI-assisted adoption of PQC in real-world software engineering.

\section{Secure Coding Drift in PQC (SCD-PQC)}
\label{SCD-PQC}

We define \textbf{SCD-PQC} as a progressive, behaviourally reinforced deviation from secure Post-Quantum Cryptography (PQC) implementation practices caused by sustained reliance on LLM-generated code within AI-mediated development workflows. This phenomenon reflects a gradual weakening of security-critical decision-making, driven by repeated interaction with AI systems that produce syntactically valid but not necessarily secure cryptographic code \cite{GLAS2026104842}. 

To illustrate, consider a developer named John implementing a lattice-based PQC key encapsulation mechanism. Initially, John carefully verifies constant-time properties and parameter selections. Over time, however, he increasingly accepts LLM-generated implementations for sampling noise vectors and performing modular arithmetic without full validation. Although the generated code appears correct and passes functional tests, it may introduce timing leakage or incorrect usage of randomness. As this pattern persists, John’s ability and motivation to independently reason about PQC security properties gradually deteriorate, resulting in a measurable shift from secure to insecure coding practices.

\subsection{SCD-PQC model}

The SCD-PQC model consists of three interacting stages as illustrated in Figure ~\ref{driftModel} :

\begin{figure}[t]
\centering
\includegraphics[width=\columnwidth] {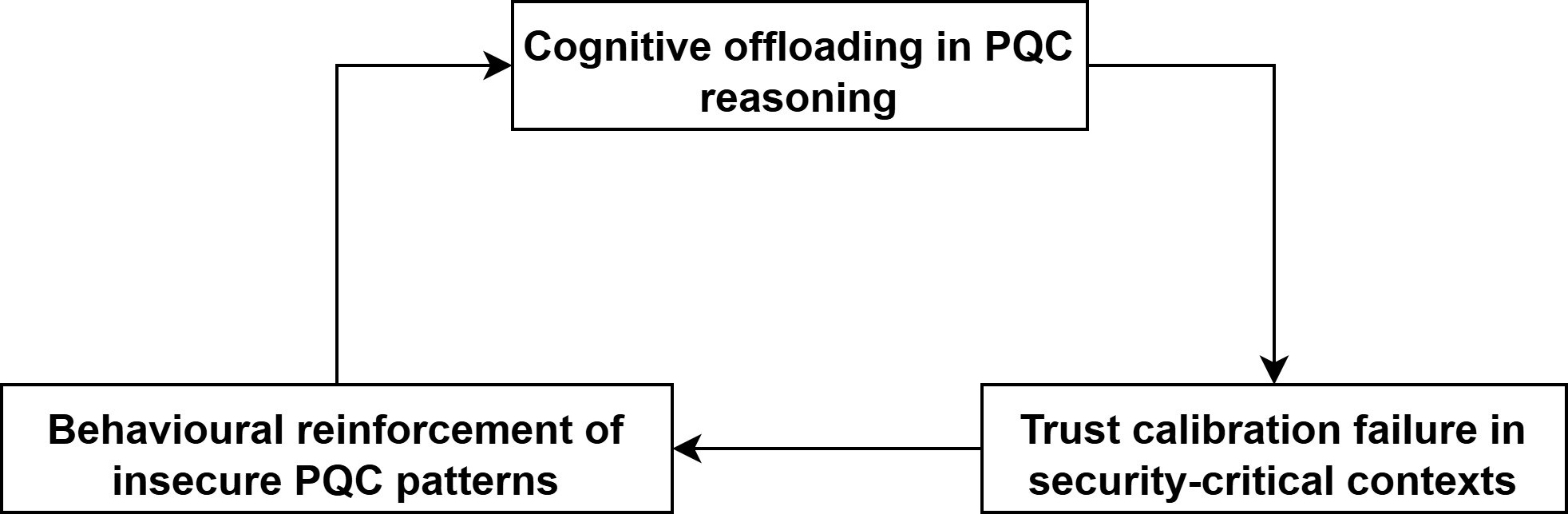}
\caption{Secure Coding Drift in PQC (SCD-PQC) Model}
\Description {Diagram showing three stages of the SCD-PQC model connected by arrows in a feedback loop.}
\label{driftModel}
\end{figure}

\begin{itemize}
    \item \textbf{Cognitive offloading in PQC reasoning:} Developers delegate complex PQC implementation tasks, such as modular arithmetic optimisations or noise sampling, to LLMs, reducing their direct involvement in security-critical logic.

    \item \textbf{Trust calibration failure in security-critical contexts:} Repeated exposure to syntactically correct and plausible outputs leads developers to overestimate the correctness of LLM-generated cryptographic code, even though it may lack guarantees for constant-time execution or side-channel resistance.

    \item \textbf{Behavioural reinforcement of insecure PQC patterns:} Acceptance and reuse of generated code reinforce insecure practices, such as improper parameter selection, unsafe randomness usage, or leakage-prone implementations. Over time, these practices become default development behaviour rather than exceptions.
\end{itemize}

\subsection{Possible Threat Vector and Impact}

The SCD-PQC model assumes an adversarial and partially untrusted LLM integrated into the software development pipeline. Within this setting, security risks emerge from three primary sources: (i) the generation of syntactically correct but insecure cryptographic code, (ii) adversarial manipulation through prompt injection or poisoned training data, and (iii) developer over-reliance that weakens independent validation of cryptographic correctness.

In this threat setting, the attacker does not require direct compromise of the target system. Instead, security degradation arises indirectly through routine developer interaction with the LLM. Even in the absence of malicious intent, repeated acceptance of AI-generated PQC code can introduce systematic vulnerabilities, including non-constant-time execution leading to timing and side-channel leakage, improper randomness generation, and incorrect or unsafe parameter selection.

The primary manifestations of this threat vector include:

\begin{itemize}
    \item Introduction of non-constant-time implementations that are vulnerable to timing and side-channel attacks
    \item Misuse of cryptographic APIs and incorrect parameter configuration
    \item Reinforcement and propagation of insecure coding patterns present in training data
    \item Targeted exploitation via adversarial prompts that manipulate cryptographic code generation
\end{itemize}

The impact of this threat vector is particularly critical in PQC-enabled software systems, where failures are often non-functional and therefore difficult to detect. Unlike traditional software defects that typically result in immediate system malfunction, PQC implementation flaws may silently weaken security guarantees while preserving apparent functional correctness \cite{hekkala_implementing_2023, bagirovs_applications_2024}. As a result, such vulnerabilities can remain latent until they are exploited under adversarial conditions.

Overall, SCD-PQC reframes the security problem from isolated implementation errors to a continuous and cumulative degradation process that simultaneously affects developer behaviour and system-level cryptographic assurance.

\section{Behavioural Amplification in AI-Native Developers}

Developers operating within AI-mediated environments prioritise speed, iteration, and convenience in their interaction patterns \cite{Insecure, AI_worst}. Although these behaviours increase productivity, they reduce opportunities for deep inspection of cryptographic implementations \cite{devDebt1,devDebt2}. In PQC contexts, where the correctness depends on subtle properties (e.g., leakage resistance), these behaviours amplify SCD-PQC \cite{toruan2026security}. Importantly, this is not a failure of developers, but a systemic shift in software engineering practices driven by AI integration \cite{iacdm2026technical}. This observation motivates interventions that target behavioural dynamics, rather than solely improving the quality of code generation.

To illustrate this behavioural shift, consider the implementation example of a PQC key encapsulation mechanism as previously described in Section~\ref{SCD-PQC}. A developer named John initially applies careful verification of constant-time properties and parameter selection. However, as development progresses within an AI-mediated workflow, he increasingly relies on LLM-generated code for tasks such as noise sampling and modular arithmetic, often without performing full security validation. Although the generated code appears correct and passes functional tests, it may still introduce timing leakage or misuse of randomness. 

A similar pattern emerges when the developer requests an LLM to generate PQC implementation code. The resulting implementation produces correct outputs under standard execution and satisfies unit tests. Due to time pressure and perceived reliability of prior AI outputs, the developer integrates the code without auditing its source or execution-time behaviour. Over successive iterations, such AI-assisted decisions accumulate across the system. While each integration appears reasonable, the overall codebase gradually incorporates non-verified PQC components, increasing exposure to side-channel leakage, non-constant-time executions, and weak randomness generation.

This example demonstrates how behavioural amplification in AI-native development reinforces the Secure Coding Drift process described in Section~\ref{SCD-PQC}, transforming isolated acts of code acceptance into a sustained degradation of PQC assurance.

\section{The Gamified Fix}

\begin{figure}[t]
\centering
\includegraphics[width=\columnwidth] {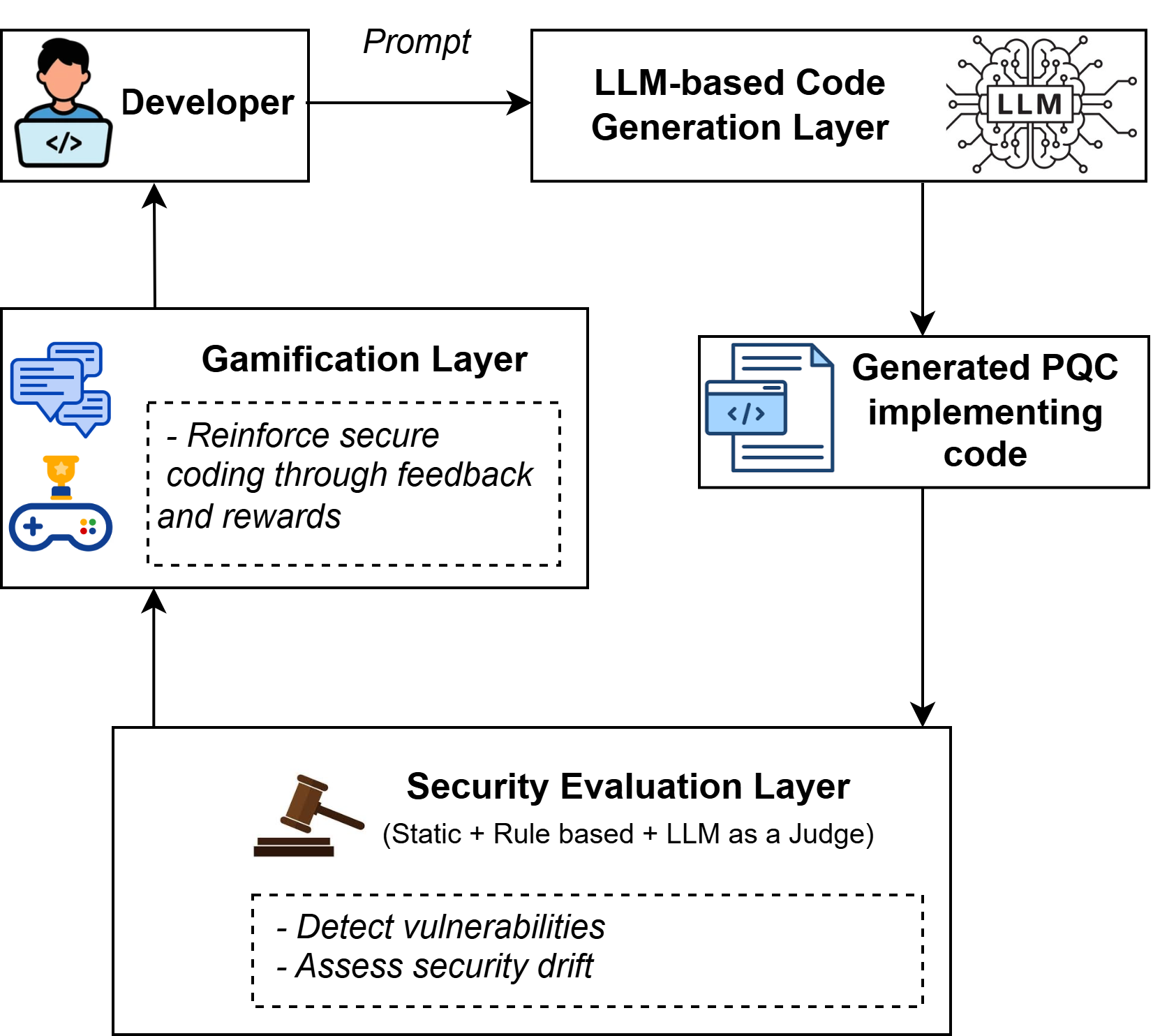}
\caption{LLM-augmented Gamified PQC engineering framework} 
\Description {Three-layer framework showing LLM-based PQC code generation, security evaluation with hybrid static and LLM-based analysis, and a gamification layer providing behavioural feedback and reinforcement in a closed-loop system.}
\label{method}
\end{figure}

We propose a Gamified LLM-augmented PQC Framework with three core components as illustrated in Figure ~\ref{method}. The operational structure of the framework is organised into three interacting layers. The interaction between these layers defines a structured execution flow that is grounded in both software security engineering principles and behavioural computing theory. 

\begin{itemize}
    \item \textbf{LLM-based code generation layer:} This layer generates PQC implementations while exposing intermediate outputs for inspection and traceability. It enhances transparency in code synthesis by enabling developers to observe and critically evaluate the generation process rather than treating outputs as final artefacts. Conceptually, this layer operationalises AI-assisted synthesis of PQC implementations, where the LLM functions as a probabilistic program generator conditioned on developer intent. This aligns with contemporary software engineering practice, where generative models are increasingly embedded into development pipelines to reduce implementation overhead while maintaining functional correctness.
\end{itemize}

\begin{itemize}
    \item \textbf{Security evaluation layer:} This layer performs automated security assessment by combining rule-based static analysis with LLM-as-a-Judge mechanisms. Generative AI tools are leveraged to support semantic reasoning about cryptographic correctness that is difficult to capture through deterministic rules alone \cite{GenAIToSSE}. This hybrid design strengthens vulnerability detection for issues such as timing leakage, insecure randomness, and cryptographic parameter misuse, while complementing traditional static verification techniques. 
    
    More specifically, the security evaluation layer introduces a hybrid verification paradigm that integrates deterministic and probabilistic reasoning. Rule-based static analysis provides formalised checks for known vulnerability classes, whereas LLM-as-a-Judge components approximate expert-level reasoning for cryptographic code review. This dual mechanism is particularly relevant in PQC settings, where vulnerabilities often arise from subtle violations of implementation constraints such as constant-time execution and parameter validity. By embedding generative AI within the evaluation loop, the framework extends conventional static analysis with context-sensitive judgment, enabling the detection of non-trivial security defects that are difficult to encode as explicit rules \cite{GenAIToSSE}. 
    
    In addition, the evaluation process produces a drift indicator, which quantifies the extent to which developer behaviour is shifting toward insecure implementation patterns based on observed vulnerabilities and repeated coding decisions.
\end{itemize}

\begin{itemize}
    \item \textbf{Gamification layer:} This layer operationalises behavioural reinforcement through scoring, feedback, progression, and challenge-based learning. It incentivises correct verification actions, rewards secure code corrections, and promotes active identification of vulnerabilities, while discouraging unvalidated acceptance of LLM-generated outputs. 
    
    From a behavioural perspective, the gamification layer is grounded in security and human-computer interaction research, which shows that sustained behavioural change is more effectively achieved through reinforcement mechanisms than through passive instruction alone. By introducing structured feedback loops, visible progression, and reward-based incentives, this layer directly shapes developer decision-making across repeated interactions. Prior studies demonstrate that such mechanisms improve engagement with secure coding practices and strengthen long-term adherence to security guidelines by reinforcing intrinsic motivation and desired behavioural patterns \cite{game, game2, game3}. Within the proposed framework, this layer functions as a behavioural control mechanism that regulates how developers respond to AI-generated code and security evaluation feedback.
\end{itemize}

Collectively, the three layers form a comprehensive feedback system in which code generation, automated evaluation, and behavioural reinforcement operate in a continuous loop. This design is consistent with cybernetic principles of adaptive systems, where feedback is used to regulate system behaviour over time. In the present context, feedback is not only technical (i.e., vulnerability detection) but also behavioural (i.e., shaping developer interaction patterns), thereby enabling simultaneous improvement of code quality and mitigation of Secure Coding Drift in PQC (SCD-PQC).

\subsection{Gamification Rationale and Behavioural Alignment}

Gamification is promising in this context because Secure Coding Drift is fundamentally a behavioural phenomenon rather than a purely technical one. In AI-native development, developers optimise for speed and correctness signals provided by LLMs, often at the expense of security reasoning. Gamification reshapes this incentive structure by introducing explicit rewards for secure behaviour, such as identifying vulnerabilities, rejecting insecure suggestions, and correctly validating cryptographic properties \cite{game, game2, game3}. This shifts developer attention from output acceptance to process quality, thereby improving trust calibration and reducing cognitive offloading.

\subsection{Metrics and Mechanisms}

The proposed framework operationalises behavioural change through measurable indicators distributed across the security evaluation and gamification layers \cite{devDebt1}. These metrics are designed to quantify both technical security quality and behavioural adaptation over time.

Within the \textbf{security evaluation layer}, the generated PQC code is analysed to compute a \textit{drift indicator}, which estimates the degree to which the developer’s coding behaviour is moving toward insecure implementation patterns. This indicator is derived from multiple security-related metrics, including vulnerability density (number of detected vulnerabilities per code segment), constant-time violations, and cryptographic misuse rate, such as unsafe randomness usage or incorrect parameter selection. Collectively, these metrics provide a technical measure of implementation quality and act as early warning signals of SCD-PQC. 

Within the \textbf{gamification layer}, evaluation outcomes are translated into behavioural feedback using scores, rewards, and challenge-based learning tasks. This layer aims to influence developer decision-making by reinforcing secure validation behaviours and discouraging blind acceptance of AI-generated outputs. This layer uses the following core metrics:

\begin{itemize}
    \item \textbf{Security correctness score:} Measures whether the generated PQC code satisfies key security requirements, including constant-time execution, correct parameter selection, and safe randomness usage.

    \item \textbf{Verification engagement index:} Measures the frequency and depth of developer validation actions, such as manual code review, invocation of static analysis tools, and adversarial testing attempts.

    \item \textbf{Drift reduction score:} Quantifies the reduction of repeated insecure coding patterns over time, indicating whether secure coding behaviour improves through repeated interaction with the framework.
\end{itemize}

Gamification mechanisms are implemented through points, levels, security scores, and challenge tasks. Developers receive higher rewards when they successfully identify subtle cryptographic flaws, reject insecure LLM-generated suggestions, or apply secure corrections. Conversely, repeated acceptance of unverified or insecure code lowers progression rates and triggers corrective challenges, thereby encouraging reflective and security-aware development behaviour.

\subsection{Design Principles}

The framework is governed by four design principles:

\begin{itemize}
    \item \textbf{Immediate Feedback:} Security-related actions are reflected in real-time scoring updates.
    \item \textbf{Cognitive Minimalism:} The system avoids excessive cognitive load on developers.
    \item \textbf{Interpretability:} All scoring signals are transparent and explainable.
    \item \textbf{Robustness to Gaming:} The reward structure discourages superficial compliance behaviours.
\end{itemize}

\subsection{Illustrative Case Study}

Consider a developer implementing a lattice-based PQC key encapsulation mechanism using the proposed framework. Initially, the LLM generates a valid implementation of a noise sampling function. The security evaluation layer flags a potential non-constant-time operation. The gamification layer assigns the developer a challenge: identify and correct the vulnerability to earn progression points. The developer modifies the implementation by introducing constant-time sampling and verifies randomness sources using static analysis tools. As a result, the system rewards secure correction rather than passive acceptance. Over repeated tasks, the developer transitions from accepting default LLM outputs to actively auditing cryptographic properties before integration.

\subsection{Proposed Evaluation Plan}

The effectiveness of the proposed framework will be evaluated through a mixed-method experimental design. We plan to conduct a controlled study involving two groups of AI-assisted developers: one group using standard LLM-based development tools and another using the proposed gamified PQC framework. Both groups will be assigned PQC implementation tasks containing intentionally embedded vulnerabilities.

The framework will be assessed using both quantitative and qualitative measures, including vulnerability detection rate, secure code acceptance rate, time-to-correction, and behavioural consistency across repeated tasks. In addition, longitudinal analysis will be performed to examine whether Secure Coding Drift decreases over time in the intervention group.

Statistical comparison between the two groups will be used to determine the effectiveness of gamification in mitigating behavioural degradation in AI-native PQC development environments. This evaluation will provide empirical evidence on whether the proposed framework improves both secure coding behaviour and overall cryptographic implementation quality.

\section{Discussion and Research Directions}

The proposed SCD-PQC model and gamified mitigation framework open several important research directions at the intersection of generative AI, PQC, software engineering, and human-centred security. More broadly, this work shifts the discussion from viewing AI-assisted secure coding as a purely technical problem toward understanding it as a socio-technical challenge involving both machine-generated vulnerabilities and human behavioural adaptation.

The framework provides a basis for studying secure coding behaviour in AI-native development environments. It enables longitudinal analysis of PQC coding drift and comparative studies between human-only, standard AI-assisted, and gamified workflows to understand how interaction models affect security outcomes.

An important direction for future work is the empirical evaluation of the proposed framework. Controlled and longitudinal studies are required to evaluate whether the framework reduces SCD-PQC. Such studies may measure trust calibration, verification frequency, and vulnerability acceptance rates. The framework can also be implemented as an IDE plugin or development tool to support real-world deployment and evaluation.

From a generative AI perspective, future research should focus on improving LLM robustness, LLM-as-a-Judge reliability, prompt injection resistance, and explainable security reasoning. Domain-specific LLMs for cryptographic programming are another promising direction.

From a PQC software engineering perspective, the work motivates new security metrics such as side-channel vulnerability density, PQC compliance scores, and cryptographic misuse rates. It also encourages integration of formal verification and symbolic analysis into AI-assisted development pipelines.

From a human behavioural perspective, future studies should investigate cognitive offloading, trust calibration, and over-reliance in AI-native developers. These factors are critical for designing interventions that sustain secure coding behaviour, especially as personalised AI tools become more prevalent.

Overall, addressing these challenges requires interdisciplinary collaboration across generative AI, PQC, software engineering, and human-centred security. Such collaboration will be critical for ensuring that future AI-assisted software development improves productivity without compromising long-term PQC security.

\section{Conclusion}

This paper introduced SCD-PQC, a behavioural vulnerability emerging from LLM integration in PQC software development. It reframes security degradation as a longitudinal human–AI interaction rather than a static coding issue, extending prior vulnerability analysis. To address this, a gamified LLM-augmented framework was proposed, incorporating continuous feedback, adversarial evaluation, and behavioural incentives into development workflows. This approach repositions LLMs as active security-aware agents rather than passive code generators. Overall, the work contributes a conceptual model and mitigation strategy for secure AI-assisted PQC development and identifies future research at the intersection of LLMs, software security, and human behaviour.

\section{Relevance to Generative IR}

Although this work is situated in secure software engineering, it is closely related to generative information retrieval systems. Modern LLM-assisted coding environments operate through a retrieval-and-generation process in which prompts, contextual information, and retrieved code examples influence generated outputs. From this perspective, secure coding drift can be viewed as a consequence of repeated interaction with generative systems, where developers increasingly rely on retrieved and generated artifacts without sufficient verification. Understanding and mitigating such behavioural effects contributes to broader efforts to improve trustworthiness, reliability, and human oversight in generative IR workflows.

\section*{Positionality Statement}

This work is positioned at the intersection of PQC, software engineering, generative AI, and human-centred security. As researchers with backgrounds in cybersecurity, secure software engineering, and human-centred security, we approach PQC migration not only as a cryptographic challenge but also as a socio-technical problem involving developer cognition, behaviour, and decision-making. Our perspective is shaped by the assumption that the security risks of AI-assisted software development cannot be fully understood by code-level vulnerability analysis alone. We argue that sustained interaction with LLMs influences how developers reason about security, calibrate trust, and perform verification. This assumption directly informs our proposal of Secure Coding Drift in PQC (SCD-PQC) as a behavioural model rather than a purely technical failure model. We also adopt a human-centred security perspective, which emphasises that insecure outcomes often emerge from the interaction between humans, tools, and workflows rather than from isolated developer mistakes. Consequently, our proposed mitigation framework prioritises behavioural interventions, particularly gamification and continuous feedback, to reinforce secure coding practices in AI-native development environments. We acknowledge that our perspective may bias the proposed framework towards behavioural interventions over organisational, economic, or regulatory solutions. Furthermore, our model is currently conceptual and has not yet been empirically validated in large-scale industrial settings. Future empirical studies involving diverse developer populations and real-world PQC implementations are necessary to evaluate and refine these assumptions.

\section*{Acknowledgement of Country}

The authors acknowledge the peoples of the Woi Wurrung and Boon Wurrung language groups of the eastern Kulin Nation on whose unceded lands ACM SIGIR 2026 is hosted. We pay our respects to their Elders past and present, and extend that respect to all Aboriginal and Torres Strait Islander peoples today and their continuing connection to land, sea, sky, and community.

\bibliographystyle{ACM-Reference-Format}
\bibliography{software}

\end{document}